# Spatial Contrast: A Vintage Retinal Model of Color Vision Produces the Center-Surround Opponent Color Phenomena; The Model Was Previously Shown To Produce Mutually Exclusive Opponent Colors by a Different Mechanism

Lane Yoder
Department of Science and Mathematics, retired
University of Hawaii, Kapiolani
Honolulu, Hawaii
lyoder@hawaii.edu
NeuralNanoNetworks.com

**Abstract**

The on-off phenomena of opponent colors in center-surround may be the best-known facts of retinal processing of information. Apparently, however, no explicit model has been proposed that shows how neurons can be connected to produce the center-surround phenomena. Here it is shown that a previous simple model of color vision can produce these phenomena, including the detection of edge orientation and motion famously discovered by Hubel and Wiesel. The model was previously shown to produce major phenomena central to color vision, including mutually exclusive opponent colors.

Although the opponencies of mutually exclusive colors and center-surround involve the same color pairs, red-green and blue-yellow, the model produces them by two different mechanisms. Perceptions of two colors are mutually exclusive because only one cone type can have the most, or least, absorption of photons. Two colors have the on-off opponency of center-surround because they have the same network designs up to the ganglion cells, with the inputs reversed there. On-off opponencies with different colors are possible, but natural selection evidently did not choose them.

## 1. Introduction

"The center-surround architecture is probably the most well-known fact about retinal filtering." [1, p. 28].

The opponent color pairs are red-green and blue-yellow. Their opponencies are expressed in mutually exclusive color perception and in a center-surround arrangement of retinal cone cells such that opponent color photostimuli have an on-off effect on a ganglion cell. This effect makes possible the properties of spatial color contrast such as detection of ripe fruit, edges, and movement.

An explicit retinal network model of color vision showed how neurons can be connected to produce major phenomena central to color vision, including mutually exclusive colors [2]. This was apparently the first network-level explanation of why the perceptions of some color pairs are mutually exclusive, and why it involves those particular colors. Technically the network is a fuzzy logic decoder that decodes the spectral information encoded in the cones' responses to a photostimulus. The fuzzy decoder model was proposed in [2], refined and extended in [3], and reviewed along with other neuronal networks in [4]. The discussion in [4], sections 3.1.1 and 4.2, may be the most accessible exposition of the model. The model will also be reviewed briefly in this article.

It was shown in [3] that each of the four colors can be produced by two similar fuzzy decoders that produce all of the color phenomena that were found in the original model (including mutually exclusive colors). That makes $2^4 = 16$ possible designs for the color vision model. These designs will also be reviewed here.

The receptive field of a retinal ganglion cell is usually defined as the region on the retina, or the region in the visual field, where light can affect the ganglion cell's response. For this article, the receptive field of a retinal ganglion cell consists of the photoreceptors that can affect its response. For a ganglion cell whose response conveys color information, the receptive field consists of cones. Most receptive fields of ganglion color cells are arranged in two regions that affect the cell in opposite ways. The regions are called "center" and "surround," but they are not necessarily concentric. If a red, green, blue, or yellow photostimulus is focused on part of one region, the ganglion cell is excited, or on. If the opponent color is focused on part of the other region, the cell is inhibited below the baseline rate, or off. Photostimuli of the other two colors have little or no effect. This is a "single-opponent" cell. For example, a ganglion cell could be center red on, surround green off.

Many opponent cells are "double-opponent." This means the cell is excited by one color in the center and inhibited by the opponent color; in the surround, the pattern is reversed. For example, a double-opponent cell could be center red on green off, surround green on red off.

Here it is shown that of the 16 possible fuzzy decoder color models, only the original model proposed in [2] can produce the center-surround on-off color phenomena. The different mechanisms for mutually exclusive colors and center-surround, and only one model that produces center-surround out of 16 that produce mutually exclusive colors, imply that mutually exclusive colors and center-surround on-off properties are two different kinds of color opponency.

The reference section of this paper is unusually short for a reason. The basic phenomena of center-surround are well known. The topic of the paper is not about the phenomena, but about how neurons can be connected to produce these phenomena. Despite the substantial amount of research on center-surround phenomena, apparently nothing else in the literature addresses this topic. Several "models" have been proposed that are based on speculation rather than explicit networks of neurons. Where connections are shown, no argument or evidence is given to show the networks would produce known results. In general the models are vague and complex. In contrast, the explicit color vision model discussed here shows that the center-surround phenomena, as well as other major phenomena central to color vision, can be generated quite simply.

**2. The neuron AND-NOT gate**

For a neuron signal consisting of spike trains, the strength of the signal is measured by the frequency of spikes. It was shown in [2] and [3] that a neuron with one excitatory and one inhibitory input with strengths X and Y, respectively, can function as a logic gate with an output signal strength that has the logic truth value X AND NOT Y. It was also shown that with access to a high input, i.e., logic value TRUE, the AND-NOT logic primitive is "functionally complete" for implementing logic operations [3]. A symbol for the AND-NOT gate is shown in Fig 1.

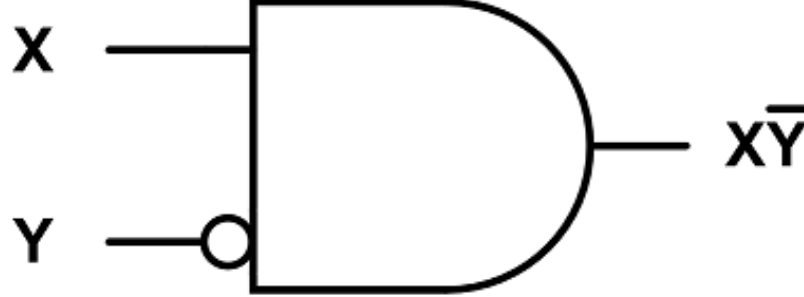

**Fig 1. A standard symbol for the logic primitive X AND NOT Y.** The rounded rectangle represents the logic AND function, and the circle represents negation (logic NOT). The AND-NOT gate can be implemented by a neuron with an excitatory input of strength X and an inhibitory input of strength Y. That is, if X and Y have binary strengths (high or low), the strength of the output signal has the logic truth value X AND NOT Y.

For non-binary inputs, i.e., inputs with strengths that are between high and low, a neuron's output is approximately the "truncated" difference between X and Y: X-Y if $X > Y$ and 0 if $X \leq Y$. The neuron's output strength is truncated because a large inhibitory input generally suppresses a smaller excitatory input. Because of neuron nonlinearities, the neuron's output strength is a measure of the truncated difference rather than the precise difference. The measure of the difference means that if $X > Y$, the response is an increasing function of X up to some maximum (more excitation increases output) and a decreasing function of Y to some minimum (more inhibition decreases output). For convenience, the truncated difference will be written as X - Y, and the truncated difference will sometimes be used as an approximation of the measure of the truncated difference.

The property that the neuron's response is a measure of the truncated difference means that the neuron AND-NOT gate can perform fuzzy logic for intermediate inputs between high and low. For this article, however, only certain minimal properties will be needed, shown in Table 1.

| |
|---|
| 1. With no input, a neuron continuously discharges at a baseline rate (low). |
| 2. If X is high and Y is low, the response is high (on). |
| 3. If X is low and Y is high, the response is hyperpolarized below the baseline rate (off). |
| 4. If X and Y are equal, there is little effect on the cell's activity. |

**Table 1. Properties of the neuron AND-NOT gate response, with excitatory input of strength X and inhibitory input of strength Y.**

## 3. A model for color vision

The original (refined) fuzzy decoder color vision model in [3] is illustrated in Fig 2.

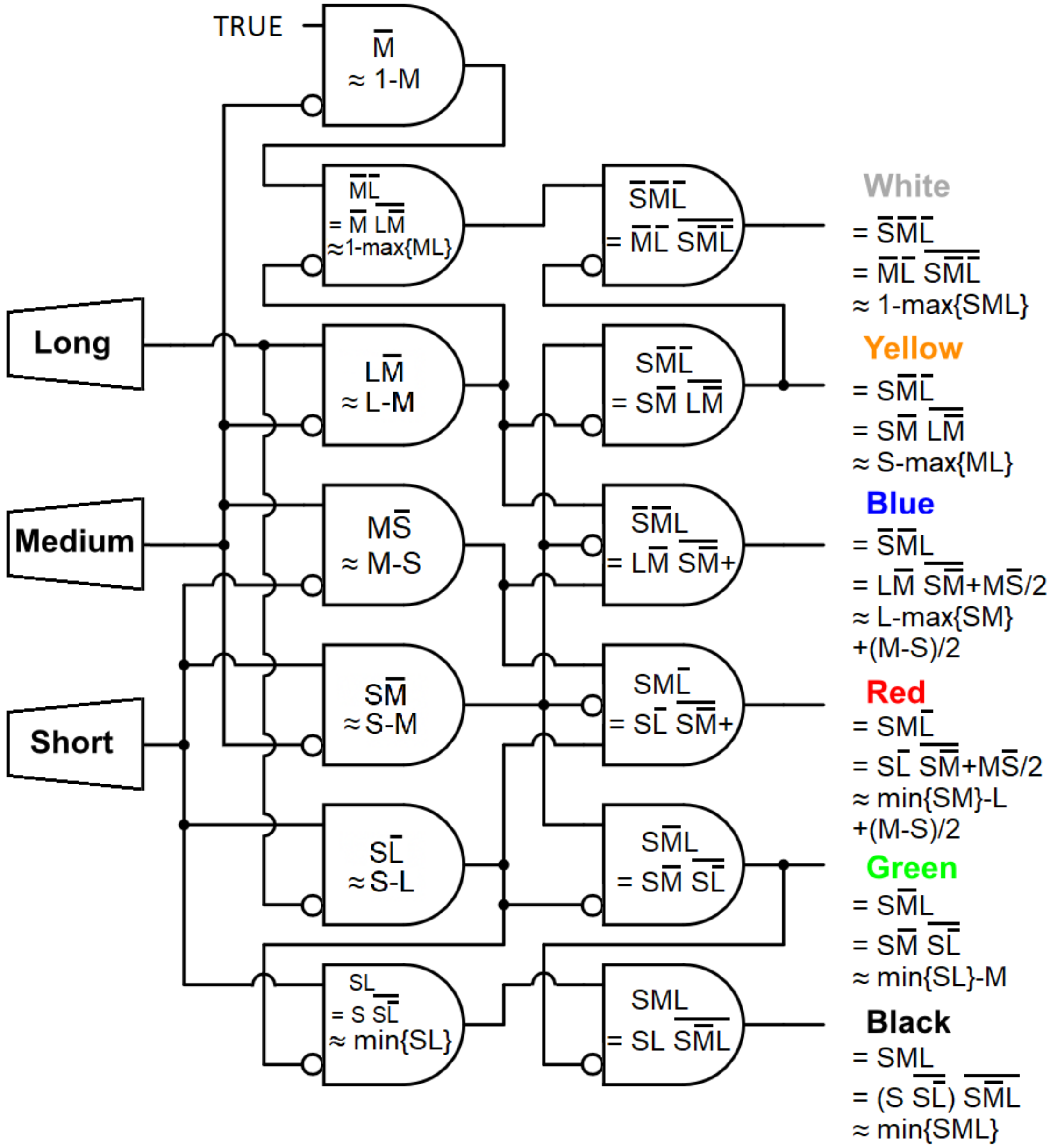

**Fig 2. Color vision model.** The network is a fuzzy logic decoder that decodes the spectral information encoded in the cones' responses to a photostimulus [4, section 4.2.1.1.]. The network receives input from three classes of spatially proximate photoreceptors that are sensitive to short, medium, and long wavelengths. The six outputs are neural correlates of the perceptions of color vision, including black and white. The strengths of the output signals are correlates of the strengths of perceptions.

The total violet and purple information ($M\overline{S}$) is conveyed through the red and blue channels. The possibilities of networks for Violet = $\overline{S}ML$ and Purple = $\overline{S}M\overline{L}$ and their possible center-surround phenomena are considered in section 6. How such networks can be included in the color vision model are presented in [4, Figs 10A and 10B].

Fig 3 shows the color networks from Fig 2.

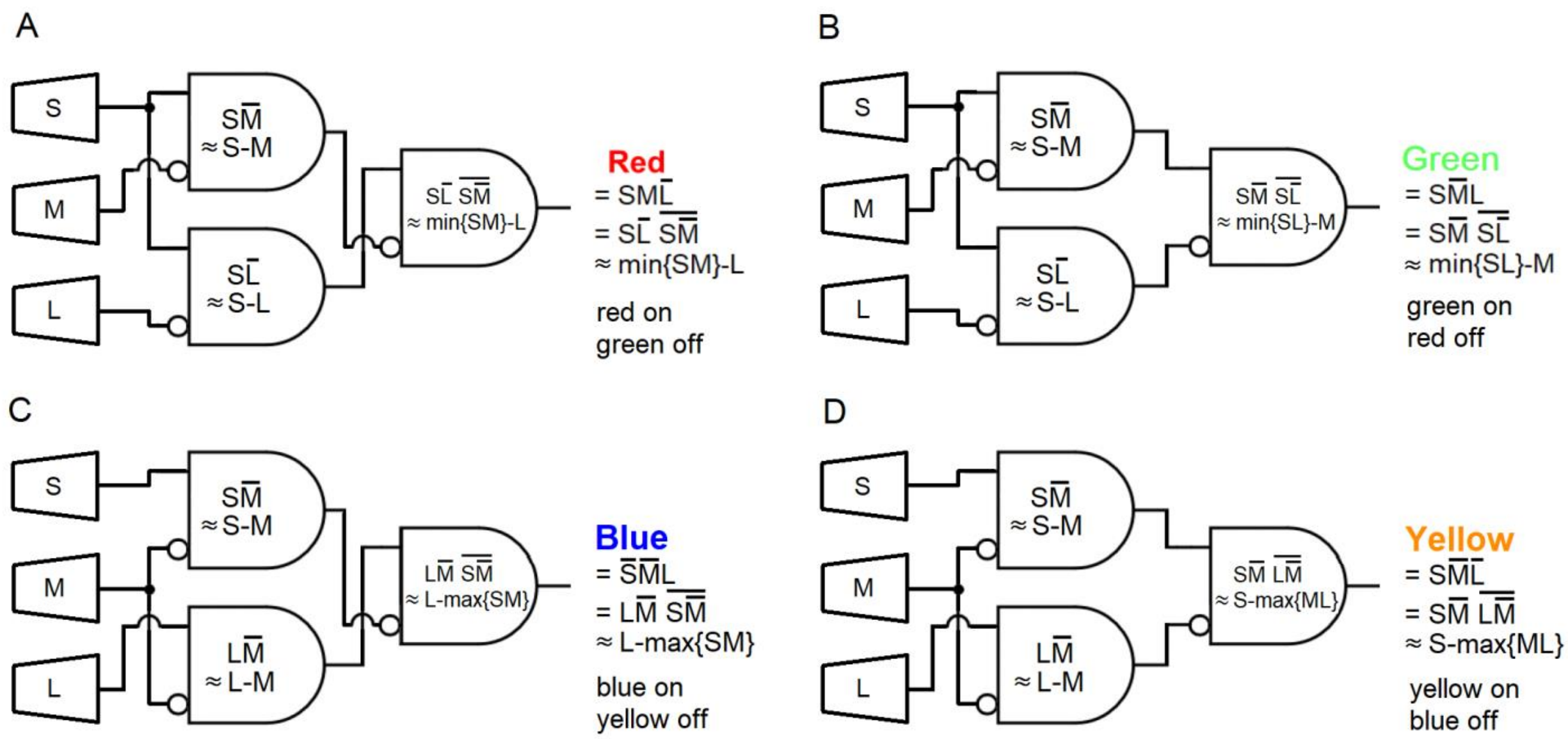

**Fig 3. The color networks from the model in Fig 2 separated into their component parts.**

The first two lines of each output in Fig 3 define its Boolean logical output. For example, Fig 3A says Red = $SM\overline{L}$, which stands for the Boolean logic statement "S AND M AND NOT L." That is, the photostimulus is perceived as red when the S and M cone outputs are high and the L cone is suppressed by the photostimulus. (Recall that vertebrate photoreceptors are spontaneously active with no stimulus and are suppressed by light of certain wavelengths.)

The third line gives the Boolean logic equivalent $S\overline{L}\ \overline{S\overline{M}}$. The equivalence follows from De Morgan's law: $S\overline{L}\ \overline{S\overline{M}} = S\overline{L}(\overline{S}$ OR $M) = SM\overline{L}$. The logic identity shows how the network can be implemented with three AND-NOT gates, as illustrated in the figure. Although there are many other possible Boolean logic identities that accomplish this implementation with AND-NOT gates, this one provides the particular fuzzy logic that agrees with color vision. The fuzzy logic result also agrees with a probability distribution in that the sum of the approximate outputs in Fig 2 is always 1.

The Boolean logic identities are used only to define how the neurons in the networks are connected. The neurons themselves produce fuzzy logic that results in neural correlates of degrees of color strength perception, as shown in the last line for each output.

The last line gives the approximate value of the Red cell's output strength, using truncated subtraction as an approximation of each cell's output: $S\overline{L}\ \overline{S\overline{M}} \approx (S-L)-(S-M)$ $= \min\{S, M\} - L$. (This final difference is also truncated.) The equation follows from considering the six possible orderings of S, M, and L. For Boolean values of 0 and 1 for S, M, and L, $\min\{S, M\} - L$ is the Boolean truth value of $SM\overline{L}$. For intermediate inputs, $\min\{S, M\} - L$ is a fuzzy logic truth value.

The Red and Green networks are the same except for the reversed inputs to the ganglion cells. The same is true of the Blue and Yellow networks. This is significant for the on-off properties of these color pairs by properties 2 and 3 of Table 1.

## 4. On-off color cell responses

Table 2 shows the Fig 3 color networks' on-off responses to color photostimuli.

| Cell ><br><br>Stimulus ∨ | S | M | L | $S\overline{M}$ | $S\overline{L}$ | $L\overline{M}$ | Red<br>$S\overline{L}\ \overline{S\overline{M}}$ | Green<br>$S\overline{M}\ \overline{S\overline{L}}$ | Blue<br>$L\overline{M}\ \overline{S\overline{M}}$ | Yellow<br>$S\overline{M}\ \overline{L\overline{M}}$ |
|---|---|---|---|---|---|---|---|---|---|---|
| red | high | high | low | **low** | <u>**high**</u> | off | **high** | <u>**off**</u> | low | low |
| green | high | low | high | <u>**high**</u> | **low** | high | <u>**off**</u> | **high** | low | low |
| blue | low | low | high | **low** | off | <u>**high**</u> | low | low | **high** | <u>**off**</u> |
| yellow | high | low | low | <u>**high**</u> | high | **low** | low | low | <u>**off**</u> | **high** |

**Table 2. The color cells' opponent (on-off) responses to color photostimuli.** The neuron responses here follow from the properties in Table 1. The main properties found here are that the Red cell is on (high response) with a red photostimulus and off (hyperpolarized) with a green photostimulus; and the Green cell is on (high response) with a green photostimulus and off (hyperpolarized) with a red photostimulus. The blue and yellow photostimuli have little or no effect on the Red and Green cells. A similar analysis shows that the Blue and Yellow cells have similar on-off responses to blue and yellow photostimuli. The bipolar cells' high inhibitory inputs that hyperpolarize the ganglion cell responses, as well as the resulting off responses, are underlined.

Importantly for center-surround, the on-off effect holds if the red photostimulus is focused only on the L cone and the green photostimulus is focused only on the M cone. The cones' overlapping sensitivity curves make the on-off effect more pronounced with the focused stimulus. This will be illustrated by example in section 5.1.2.2. This property of center-surround is important for spatial color contrast when the center and surround receive different photostimuli.

# 5. Center-surround cells

## 5.1. Single-opponent cells

### 5.1.1. Single-opponent networks

The center-surround phenomena can be created by selection of a ganglion cell's receptive field according to the on-off properties of the color cells in Fig 3. Fig 4 shows how this can be done for Red and Green ganglion cells. Blue and Yellow are similar.

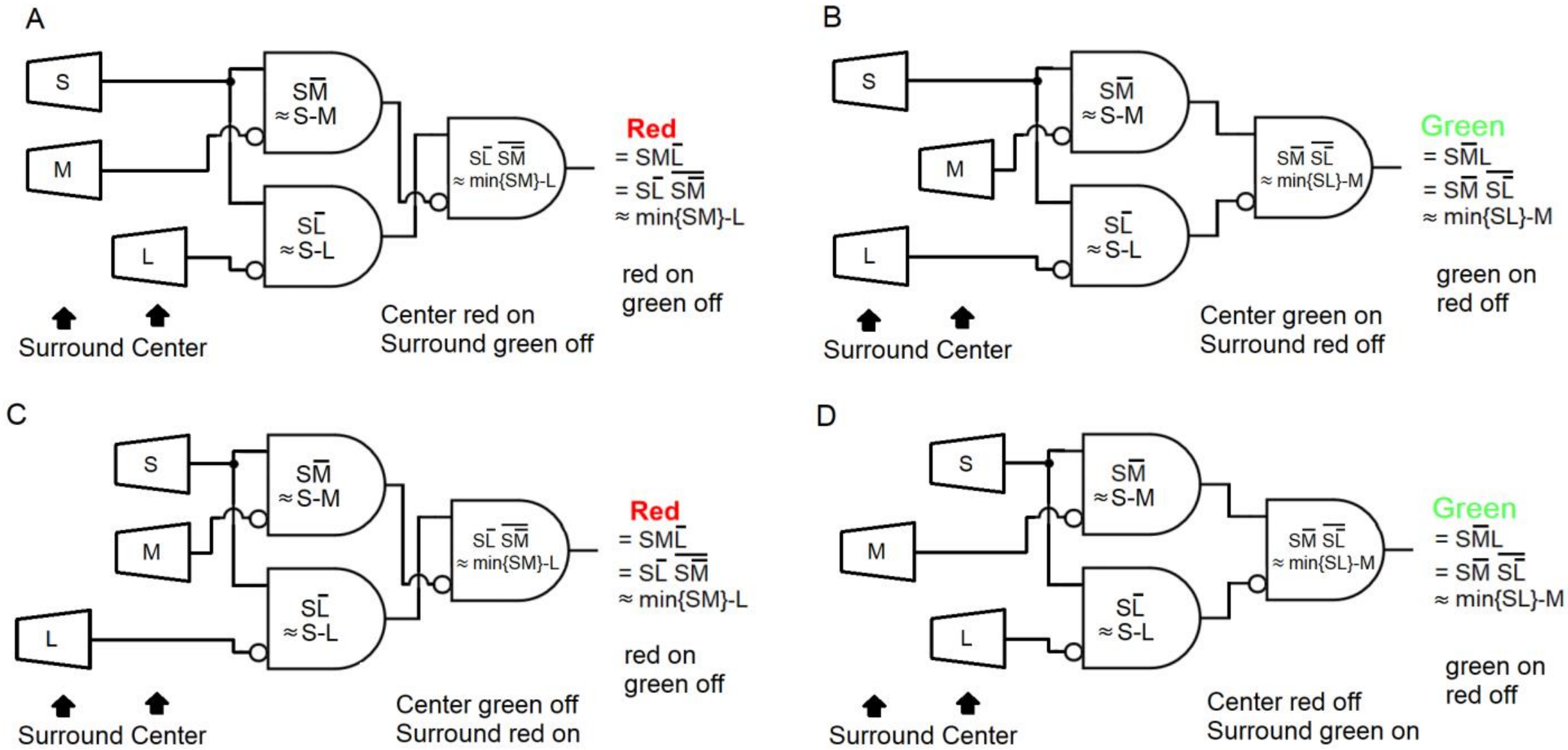

**Fig 4. Center-surround networks for Red and Green ganglion cells.** The networks in Fig 3 are shown here with the inputs separated into center and surround to produce the center-surround phenomena. The network architectures are unchanged from Fig 3; the only changes are the locations of the cones in the receptive field. Therefore the ganglion cells' on-off responses to single color photostimuli on the whole receptive field are unchanged from the Fig 3 responses shown in Table 2.

Examples of center-surround receptive fields are given in Fig 5.

**Fig 5. Example center-surround receptive fields.** The figure shows how center-surround networks of Fig 4 can be formed by the selection of cells for the receptive field. **A.** Selected cells of a receptive field for center green on. **B.** Overlapping receptive fields for center green on (left) and center red on (right). Each cone can be part of several ganglion cells' receptive fields.

Each bipolar cell's inputs from the cones could be weighted so that the inputs from each cone class effectively act as a single cone, as in Fig 4.

### 5.1.2. On-off examples

Experimental evidence shows that the on-off effect is more pronounced when the photostimulus falls only on the center or surround. The fuzzy decoder model produces this effect. Two examples illustrate the property with the Fig 4A center red on surround green off network. Truncated differences are underlined.

These photostimuli are used in the examples:

Red photostimulus: S = 0.9, M = 0.7, L = 0.1.

Green photostimulus: S = 0.9, M = 0.1, L = 0.7.

### 5.1.2.1. Photostimulus on the whole receptive field

A photostimulus on the whole receptive field necessarily has the same effect as the receptive fields with no center-surround in Fig 3.

Red photostimulus S = 0.9, M = 0.7, L = 0.1:

Red = $SM\overline{L} = S\overline{L}\ \overline{S\overline{M}} \approx$ (S-L)-(S-M) = (0.9-0.1)-(0.9-0.7) = 0.8-0.2 = 0.6. (On).

Green photostimulus S = 0.9, M = 0.1, L = 0.7:

Red = $SM\overline{L} = S\overline{L}\ \overline{S\overline{M}} \approx$ (S-L)-(S-M) = (0.9-0.7)-(0.9-0.1) = $\underline{0.2\text{-}0.8}$ = 0. (Off. The truncated difference -0.6 indicates the response is somewhat hyperpolarized by property 3 of Table 1.)

### 5.1.2.2. More pronounced on-off effect with photostimulus only on center or surround

Experimental evidence shows that the on-off effect is greater if the photostimulus falls only on the center or surround. This property is important for spatial color contrast. The decoder model produces this effect.

The same red photostimulus S = 0.9, M = 0.7, L = 0.1 focused on the center (L only):

Red = $SM\overline{L} = S\overline{L}\ \overline{S\overline{M}} \approx$ (S-L)-(S-M) = (1-0.1)-(1-1) = 0.9-0 = 0.9 > 0.6. (A greater on response.)

The green photostimulus S = 0.9, M = 0.1, L = 0.7 focused on the surround: (S&M only):

Red = $SM\overline{L} = S\overline{L}\ \overline{S\overline{M}} \approx$ (S-L)-(S-M) = ($\underline{0.9\text{-}1}$)-(0.9-0.1) = $\underline{0\text{-}0.8}$ = 0. (-0.8 < -0.6 indicates a greater off response.)

## 5.2. Double-opponent cells

### 5.2.1. A double-opponent network

A double-opponent network is shown in Fig 6.

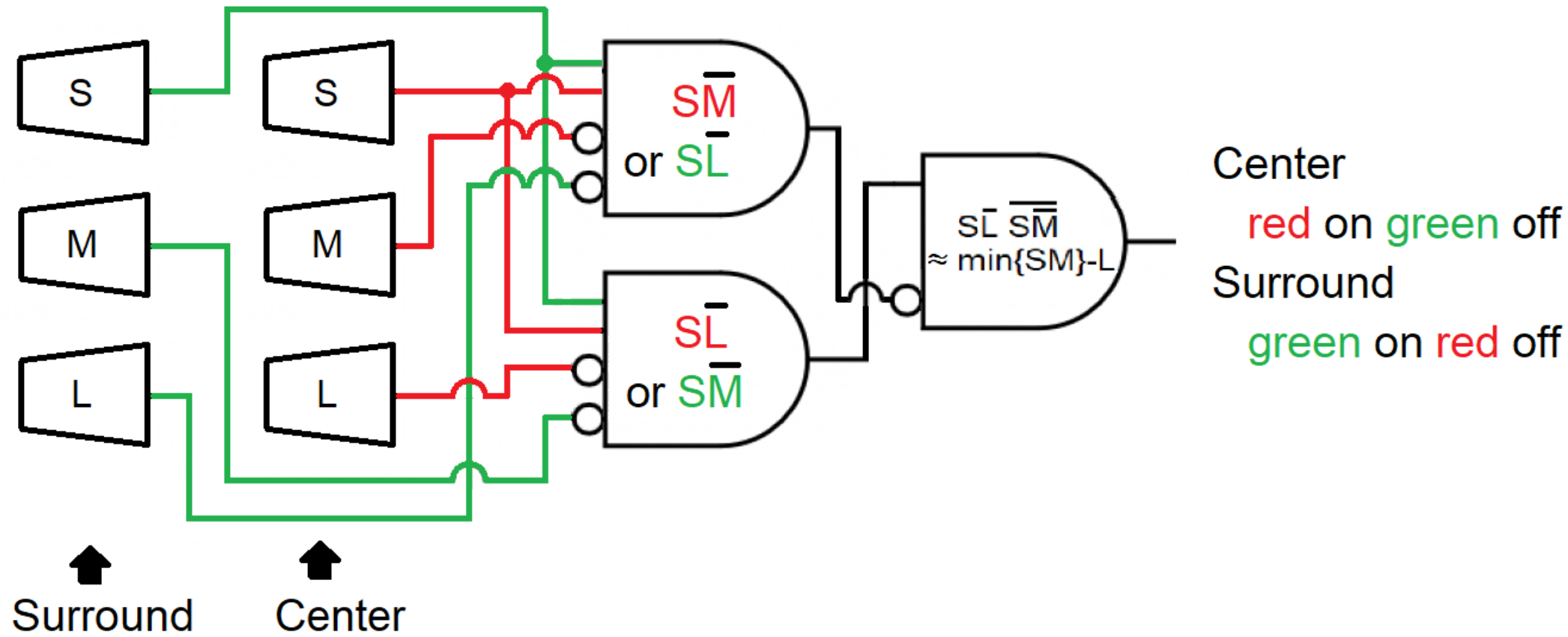

**Fig 6. A double-opponent cell that is center red on green off and surround green on red off.** Both center and surround have all three cone types. The center is connected like the red network of Fig 3A, so the center is red on green off. The surround is connected like the green network of Fig 3B, so the surround is green on red off. Reversing the two inputs to the ganglion cell makes the ganglion cell center green on red off and surround red on green off.

### 5.2.2. Examples of double-opponent spatial contrast capabilities

The four examples show that double-opponency is especially effective in detecting local color contrast. The red graphs show the responses of the center red on cell in Fig 6. The green graphs show the responses of a center green on cell.

#### 5.2.2.1. Edge detection

Fig 7 shows double-opponent edge detection.

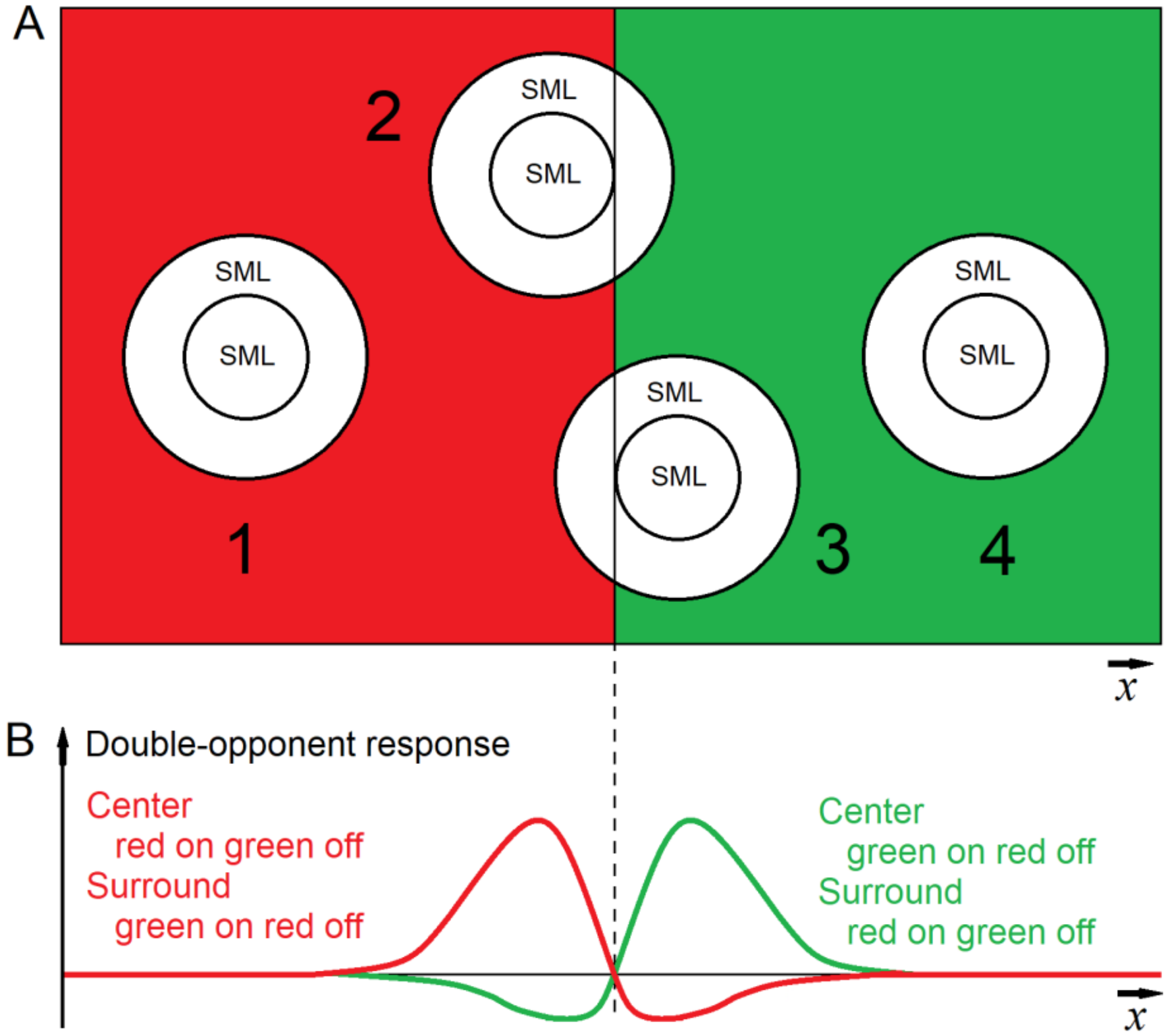

**Fig 7. Edge detection by a double-opponent cell.** Both center and surround have all three cone types as in Fig 6. **A.** A red-green image in various positions on a double-opponent cell's receptive field. **B.** Graphs of estimated responses of double-opponent cells. The red graph represents the response of the center red on cell of Fig 6. The green graph represents the response of a center green on cell.

For the red graph in Fig 7B: At position 1 in Fig 7A, the center red on and surround red off cancel each other. (This is by property 4 in Table 1. Both ganglion cells are high $S\bar{L}$ by property 2 of Table 1.) Similarly at position 4, the center green off and surround green on cancel. At position 2, the center red on is the same as position 1, but the surround red off is less than in position 1 and there is some surround green on. Both of these changes increase the ganglion cell's response. At position 3, the center is fully green off, and the surround is partially red off and partially green on. The result is some hyperpolarization below the baseline. The argument for the green graph is similar.

#### 5.2.2.2. Ripe fruit detection

Fig 8 shows the detection of a red spot on green.

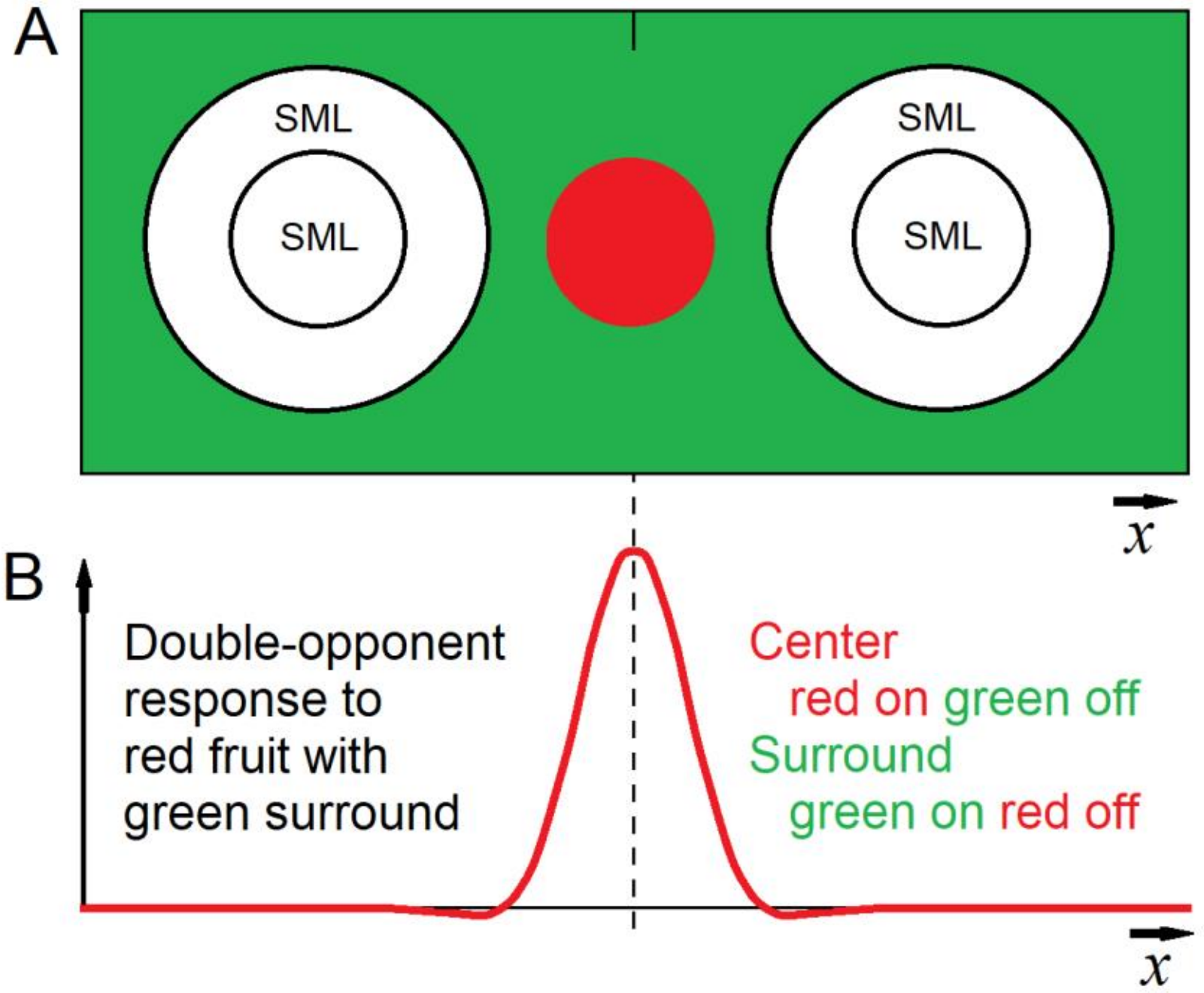

**Fig 8. Response of a double-opponent cell to red fruit surrounded by green foliage.** The cell is center red on green off and surround green on red off as in Fig 6. When the image of the fruit lies outside the receptive field, the center green off suppresses the surround green on. When the image of the fruit aligns with the receptive field center, the response is center red on and surround green on, giving a strong on response.

#### 5.2.2.3. Side-by-side on-off regions

As mentioned in the introduction, the on-off regions of the "center-surround" receptive field are not necessarily concentric.

##### 5.2.2.3.1. Side-by-side on-off regions found experimentally

Fig 9 shows the responses of side-by-side on-off regions found experimentally.

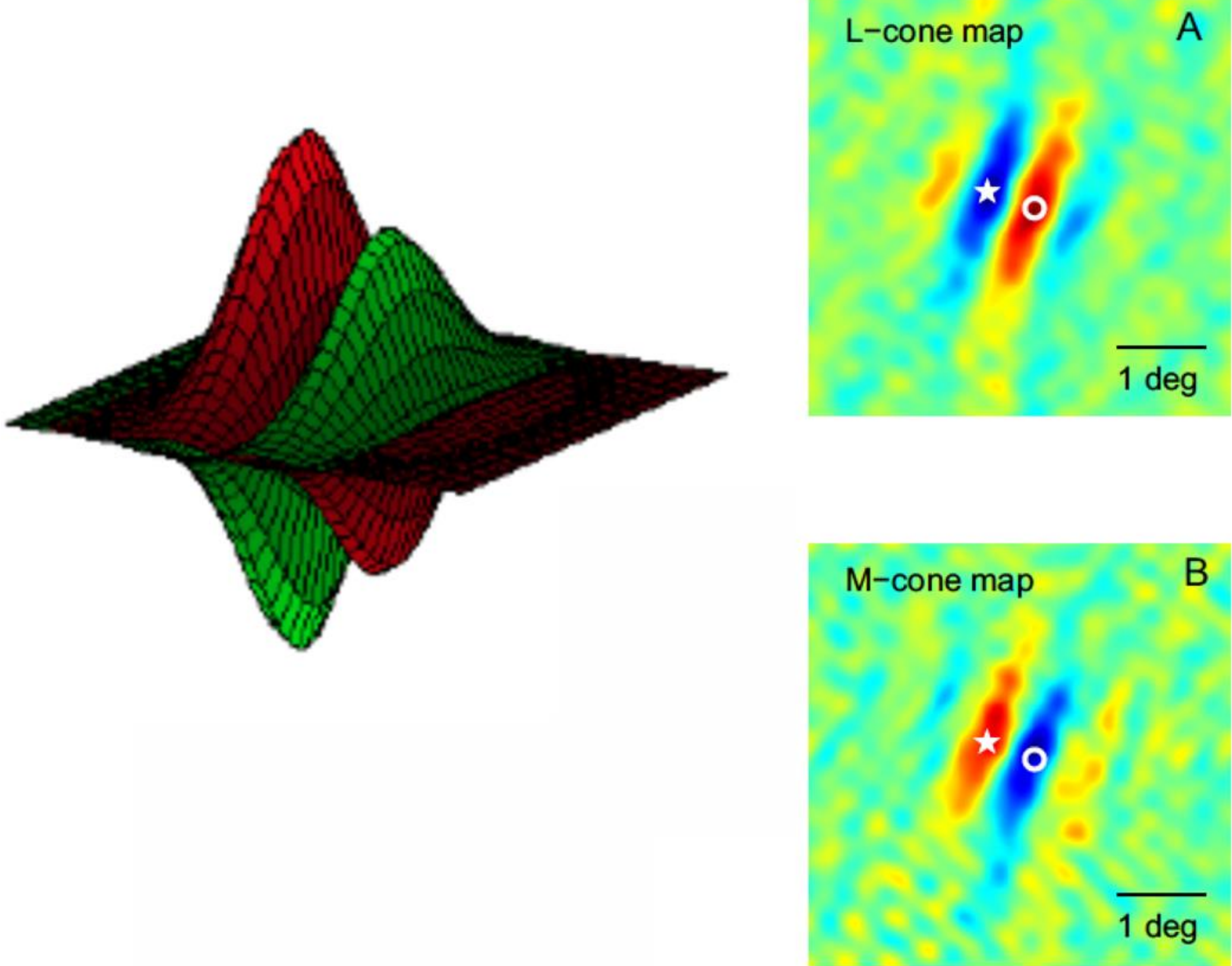

**Fig 9. Upper left red on green off, lower right green on red off** (From [5, Fig 6]). The 3D schematic graph shows a ganglion cell's approximate responses to local retinal photostimuli. The graph matches the responses of the double-opponent cell of Fig 6, with the off responses displayed as negative values. Figs **A** and **B** show the side-by-side on-off regions of the cell's receptive field.

### 5.2.2.3.2. Oriented edge detection

Fig 10 shows edge detection by a double-opponent cell with a side-by-side receptive field.

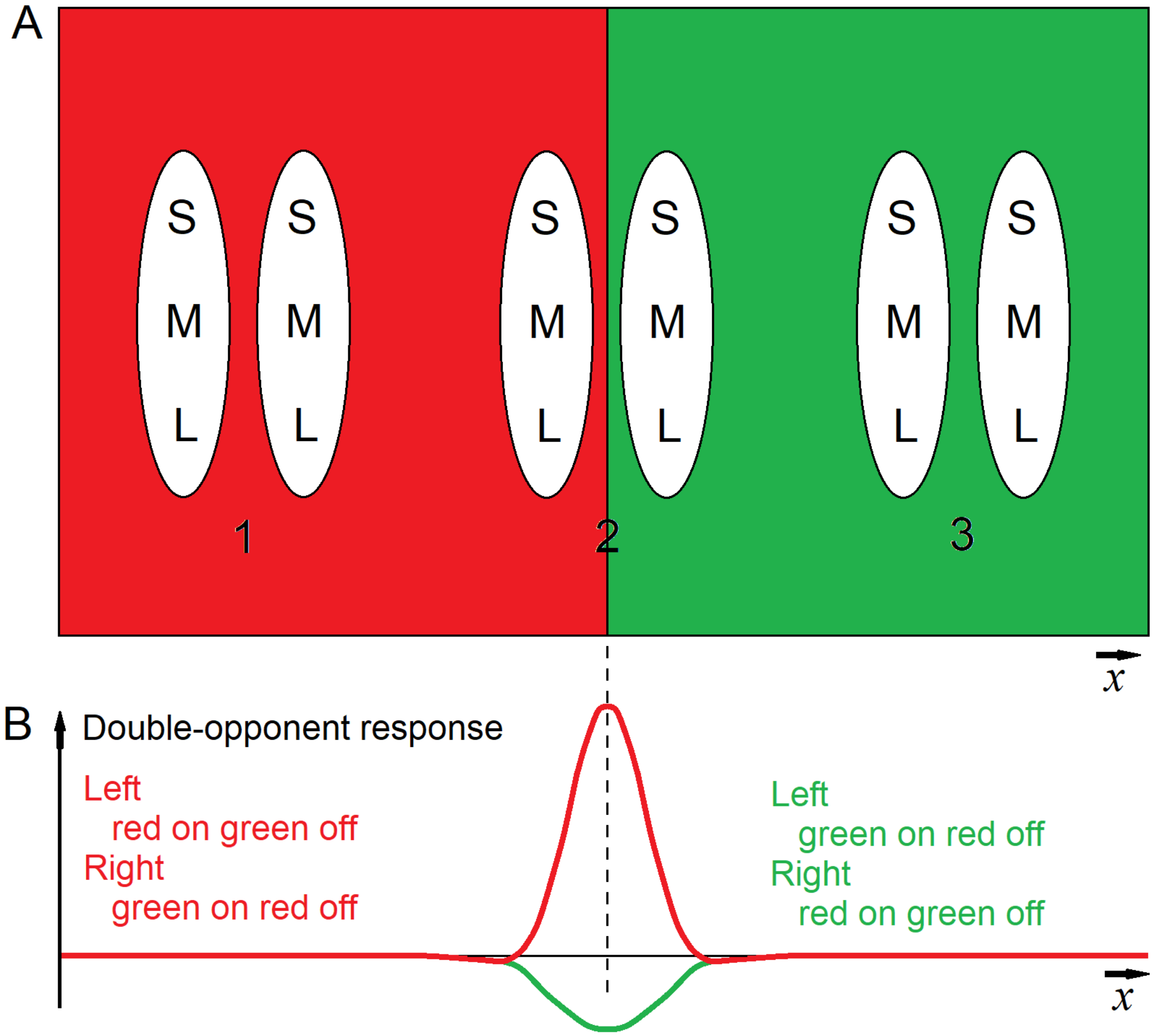

**Fig 10. Edge orientation with nonconcentric center-surround.** The figure shows how the receptive field composed of side-by-side regions can respond to an edge with a specific orientation. **A.** Color images on various positions of side-by-side on-off regions of a receptive field. **B.** The red graph represents the response of the center (or left) red on cell of Fig 6. The green graph represents the response of a center green on cell. For the red graph, at position 1, the left red on and right red off cancel each other. Similarly at position 3, the left green off and right green on cancel. At position 2, the left red on and right green on give a strong on response. Similarly, for the green graph, at position 1 the left red off and right red on cancel, and at position 3 the left green on and right green off cancel. At position 2, the left red off and right green off give a strong off response.

### 5.2.2.3.3. Directional movement detection

Fig 11 shows directional movement detection, similar to that discovered by Hubel and Wiesel [7].

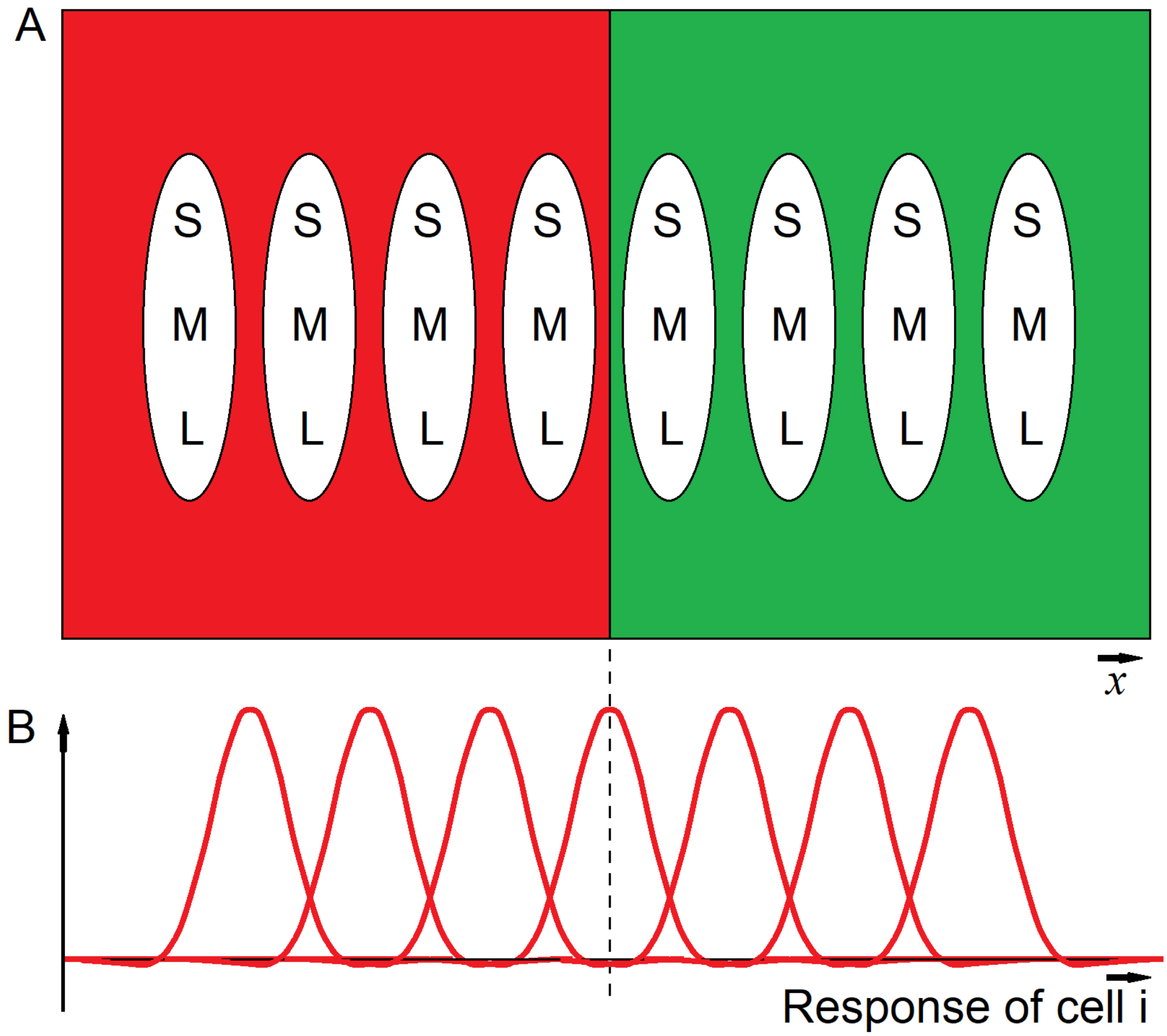

**Fig 11. Oriented edge movement detection by multiple double-opponent cells, one for each pair of side-by-side retinal regions.** As the image moves across the retina (**A**), each ganglion cell fires in sequence, giving the perception of movement (**B**). The cell responses show the location of movement in the visual field, and its speed and direction in either of the two

directions perpendicular to the axis between the two side-by-side regions. The edge motion could be a moving two-colored object or an object of one color moving across an object of the other color.

### 5.3. Shared receptive fields and bipolar cells.

Fig 12 shows how the network designs allow efficient use of cells.

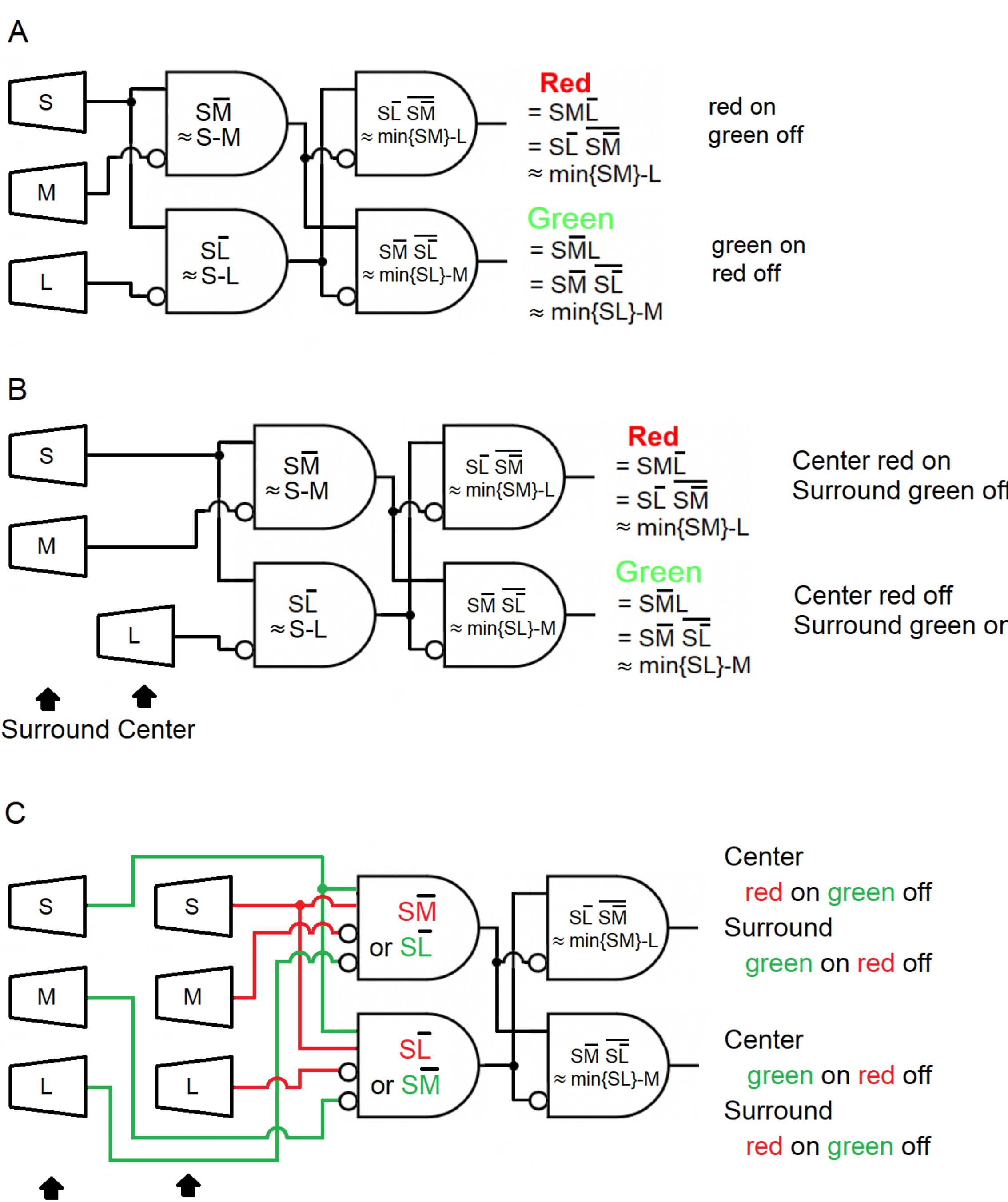

**Fig 12. Shared receptive fields and bipolar cells.** Because the Fig 3 network designs for each opponent color pair are identical except for the reversed inputs to the ganglion cells, two ganglion cells can share the same receptive field and bipolar cells. **A.** Single-opponent cells. **B.** Single-opponent center-surround cells. **C.** Double-opponent center-surround cells.

## 6. Different mechanisms produce mutual exclusion and center-surround

Several possible fuzzy decoders produce all of the phenomena of color vision found in the original model in [2, 3], including mutually exclusive colors. Only one of these models produces the known on-off properties of center-surround. This is because the fuzzy decoder model produces mutually exclusive colors and center-surround on-off phenomena by different mechanisms.

### 6.1. Two possible fuzzy decoders for each color

The set of recursive logic identities in Table 3 extended the fuzzy decoder in Fig 2 to allow any number of inputs [3].

| Recursive and reductive logic identities for a fuzzy decoder composed of AND-NOT gates | Interval measured by the decoder response | Approximate value of the decoder response |
| --- | --- | --- |
| 1. $\Pi_{i=1}^{M} X_i \Pi_{j=1}^{N} \overline{Y_j} = \{\Pi_{i=1}^{M-1} X_i \Pi_{j=1}^{N} \overline{Y_j}\}\overline{\{\overline{X_M}\Pi_{i=1}^{M-1} X_i \Pi_{j=1}^{N-1}\overline{Y_j}\}}$ | $[\max_{j=1}^{N}\{Y_j\},\ \min_{i=1}^{M}\{X_i\}]$ | $\min_{i=1}^{M}\{X_i\}-\max_{j=1}^{N}\{Y_j\}$ |
| 2. $\Pi_{i=1}^{M} X_i \Pi_{j=1}^{N} \overline{Y_j} = \{\Pi_{i=1}^{M} X_i \Pi_{j=1}^{N-1} \overline{Y_j}\}\overline{\{Y_N \Pi_{i=1}^{M-1} X_i \Pi_{j=1}^{N-1}\overline{Y_j}\}}$ | | |
| 3. $\Pi_{i=1}^{N} X_i = \{\Pi_{i=1}^{N-1} X_i\}\overline{\{\overline{X_N}\Pi_{i=1}^{N-1} X_i\}}$ | $[0,\ \min_{i=1}^{N}\{X_i\}]$ | $\min_{i=1}^{N}\{X_i\}$ |
| 4. $\Pi_{j=1}^{N} \overline{Y_j} = \{\Pi_{j=1}^{N-1}\overline{Y_j}\}\overline{\{Y_N \Pi_{j=1}^{N-1}\overline{Y_j}\}}$ | $[\max_{j=1}^{N}\{Y_j\},\ 1]$ | $1-\max_{j=1}^{N}\{Y_j\}$ |

**Table 3. Boolean logic identities that define a fuzzy decoder [3].** The recursive and reductive logic identities in the first column show how fuzzy decoders with any number of inputs can be implemented with AND-NOT gates. The second column shows the interval measured by the network's response for decoders composed of neurons. The truncated difference in the third column shows that the value of the response is approximately the length of the interval if the response of each AND-NOT gate is approximately the truncated difference of the inputs. Simpler Boolean logic identities can show how decoders can be implemented with AND-NOT gates, but the identities in the table produce the fuzzy logic interval measures in the last column for decoders composed of neurons.

The logic identities in the first column of Table 3 equate every product to a product A AND NOT B. Equations 1 and 3 show how to add one input X to A, and equations 2 and 4 add one input $\overline{\mathrm{Y}}$ to A. For example, if M = 2 and N = 1, identity 1 shows how to add X2 to $\mathrm{X1}\overline{\mathrm{Y1}}$:

$\mathrm{X1X2}\overline{\mathrm{Y1}} = \mathrm{X1}\overline{\mathrm{Y1}}\;\overline{\overline{\mathrm{X2}}\mathrm{X1}}$.

For cone responses S, M, and L, if X1 = S, X2 = M, and Y1 = L, this identity is the Red network in Fig 3A:

$\mathrm{Red} = \mathrm{SM}\overline{\mathrm{L}} = \mathrm{S}\overline{\mathrm{L}}\;\overline{\mathrm{S}\overline{\mathrm{M}}}$.

The logic identities in Table 3 imply fuzzy decoders can be implemented in more than one way. If X1 = M, X2 = S, and Y1 = L, the above identity gives

$\mathrm{Red} = \mathrm{SM}\overline{\mathrm{L}} = \mathrm{M}\overline{\mathrm{L}}\;\overline{\mathrm{M}\overline{\mathrm{S}}}$.

The bidirected graphs in Fig 13 show all possible color decoders that can be derived from Table 3. There are two possible decoders for each color.

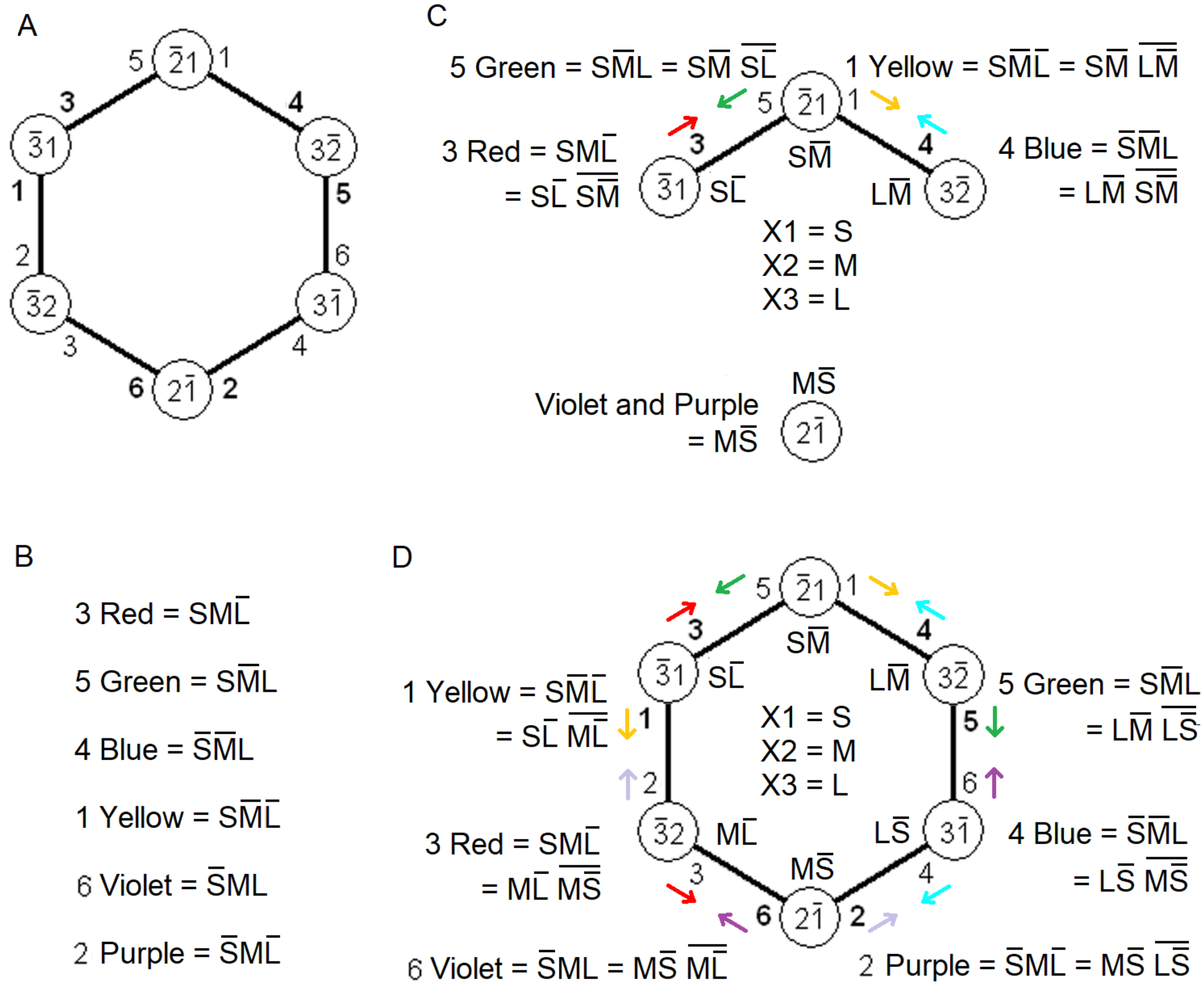

**Fig 13. All possible color decoders (adapted from [4, Fig 10]).** The graphs show how color decoders can be formed from Table 3. **A.** All of the ways a three-input decoder can be formed according to the logic identities 1 and 2 of Table 3 [6, Fig 5A], as described below. **B.** The Boolean logic identities for six possible colors. **C**. Part of the graph in A, showing the color vision model of Figs 2 and 3. **D.** The graph in A, showing the other possible ways to implement the four color cells plus Violet and Purple.

The nodes in the graphs in Fig 13 represent neurons, and their labels are shorthand for the cell's logic output. For example, in the node at the top of Fig 13A, the label $\bar{2}1$ stands for $X1\overline{X2}$ (X1 AND NOT X2). That is, the cell has excitatory input X1 and inhibitory input X2.

Two nodes are linked if the cells they represent can be connected by an AND-NOT function according to equation 1 or 2 of Table 1. Along each edge linking two nodes N1 and N2, the number next to N1 represents the decoder output $N1\overline{N2}$. For example, the number 3 in Fig 13A next to node $\overline{3}1$ represents the decoder output $X1\overline{X3}\,\overline{X1\overline{X2}} = \overline{X3}X2X1$. The number 3 is used to represent this output because the Boolean output $\overline{X3}X2X1 = 1$ when the Boolean inputs are X3X2X1 = 011, which is the binary representation of the number 3.

For each of the four colors in the original color vision model in Figs 2 and 3, either of the two possible ways of producing it shown in Figs 13C and D can be used to produce all of the color phenomena found in the original model, including mutually exclusive colors, because both ways produce the same interval measure property by Table 3. That means there are $2^4 = 16$ possible designs for the model. Fig 3 and Fig 13C show one of the 16 possible designs.

### 6.2. The two decoders' results are computationally different

The two ways of producing each color give measures of the same interval by Table 3. For example, Red = $SM\overline{L}$ is the measure of [L, min{S,M}]. The two ways of achieving this result by Fig 13D, however, are computationally different.

With the previous example of a red-yellow photostimulus: S = 0.9, M = 0.7, and L = 0.1:

$$Red_1 = SM\overline{L} = S\overline{L}\,\overline{S\overline{M}} \approx (S\text{-}L)\text{-}(S\text{-}M) = (0.9\text{-}0.1)\text{-}(0.9\text{-}0.7) = 0.8\text{-}0.2 = 0.6.$$

$$Red_2 = SM\overline{L} = M\overline{L}\,\overline{M\overline{S}} \approx (M\text{-}L)\text{-}(M\text{-}S) = (0.7\text{-}0.1)\text{-}(\underline{0.7\text{-}0.9}) = 0.6\text{-}0 = 0.6.$$

### 6.3. $Red_1$ is red on green off, and $Red_2$ is red on violet off

The green photostimulus S = 0.9, M = 0.1, and L = 0.7 hyperpolarizes the $Red_1$ cell but not the $Red_2$ cell:

Red$_1$ = $SM\overline{L} = S\overline{L}\,\overline{S\overline{M}} \approx (S-L)-(S-M) = (0.9-0.7)-(0.9-0.1) = \underline{0.2-0.8} = 0$. (off, -0.6).

Red$_2$ = $SM\overline{L} = M\overline{L}\,\overline{M\overline{S}} \approx (M-L)-(M-S) = (\underline{0.1-0.7})-(\underline{0.1-0.9}) = 0-0 = 0$. (not hyperpolarized).

The violet photostimulus S = 0.1, M = 0.8, and L = 0.9 does hyperpolarize the Red$_2$ cell:

Red$_2$ = $SM\overline{L} = M\overline{L}\,\overline{M\overline{S}} \approx (M-L)-(M-S) = (\underline{0.8-0.9})-(0.8-0.1) = \underline{0-0.7} = 0$. (-.07, hyperpolarized).

**6.4. Only one decoder model for center-surround on-off color opponency**

Of the 16 possible color vision designs from Table 3, only one way, shown in Figs 3 and 13C, provides the known opponent on-off responses of center-surround. Fig 13D shows that Red and Green could produce on-off opponency with Violet, and Blue and Yellow with Purple. But evidently natural selection chose not to do so [2], possibly because the full Violet and Purple information were not worth the cost of dedicated channels to the brain [2].

The networks for each color shown in Fig 13C were originally selected in [2] for the four colors in the color vision models in Figs 2 and 3 because only three bipolar cells are needed ($S\overline{L}$, $S\overline{M}$, and $L\overline{M}$). Minimizing material requirements may have influenced natural selection, but the on-off advantages of center-surround likely had a greater effect.

**6.5. Sixteen models can produce mutually exclusive colors by a different mechanism**

The opponency of mutually exclusive colors involves the same color pairs as center-surround opponency. But the two opponencies are produced by different mechanisms.

All 16 of the possible fuzzy decoder models of color vision produce the mutually exclusive color phenomena. Both versions of each color in Fig 13 were derived from Table 3, so they satisfy the interval measure property of the Table. For both versions of Red = $SM\overline{L}$, the value of $SM\overline{L}$ is a measure of the interval [L, min{S,M}]. This value is positive when L is the most suppressed by the photostimulus. Similarly, Green = $S\overline{M}L$ is a measure of the interval [M, min{S,L}]. This value is positive when M is the most suppressed. The L and M cones cannot both be the most suppressed, so Red and Green cannot be positive simultaneously. Blue = $\overline{S}\overline{M}L$ is a measure of [max{S,M}, L], which is positive when L is the least suppressed. Yellow = $S\overline{M}\overline{L}$ is a measure of [max{M,L}, S], which is positive when S is the least suppressed. They cannot both be least.

### 6.6. All possible opponencies

For completeness, listed here are all possible opponencies with the fuzzy decoder model if violet and purple are implemented according to Fig 13D. The violet cell has a positive response when S cones have the greatest absorption of photons. Red, green, and violet are mutually exclusive because only one of the cone types can have the greatest absorption. The purple cell has a positive response when M cones have the least absorption of photons. Blue, yellow, and purple are mutually exclusive because only one of the cone types can have the least absorption of photons. Besides red-green and blue-yellow on-off center-surround opponencies, other possibilities are red-violet, green-violet, blue-purple, and yellow-purple.

## 7. Conclusions

The center-surround models in Figs 4 and 6 were derived directly from the color vision model first proposed in [2] and shown in Figs 2 and 3. These explicit networks show how neurons can be connected to produce center-surround phenomena. This is apparently the only explicit explanation of center-surround phenomena that has been proposed.

Of the 16 possible fuzzy decoder models of color vision that follow from logic identities 1 and 2 of Table 3, and are shown in Fig 13, only the original model of Figs 3 and 13C produces center-surround color phenomena. This supports the original model as being the correct model of color vision.

The different mechanisms for mutually exclusive colors and center-surround, and only one model that produces center-surround out of 16 that produce mutually exclusive colors, imply that mutually exclusive colors and center-surround on-off properties are two different kinds of color opponency.

## 8. Acknowledgements

Figures were created with CircuitLab and MS Paint. I would like to thank David Burress for his support and many helpful suggestions and especially Ernest Greene for his strong support and for pointing out the importance of center-surround and for challenging many concepts.